\documentclass[a4paper,english,aps,notitlepage,nofootinbib,tightenlines,11pt]{revtex4-1}
\usepackage[T1]{fontenc}
\usepackage[latin9]{inputenc}
\usepackage{color}
\usepackage{amstext}
\usepackage{amssymb}
\usepackage{graphicx}
\usepackage{esint}
\makeatletter

\pdfpageheight\paperheight
\pdfpagewidth\paperwidth

\providecommand{\tabularnewline}{\\}

\pdfoutput=1
\usepackage[colorlinks=true,citecolor=blue,linkcolor=blue]{hyperref}
\usepackage{slashed}

\makeatother

\usepackage{babel}
\begin{document}

\title{Loop-enhanced rate of neutrinoless double beta decay}

\author{Werner Rodejohann and Xun-Jie Xu}

\affiliation{\textcolor{black}{Max-Planck-Institut f\"ur Kernphysik, Postfach
103980, D-69029 Heidelberg, Germany}}

\begin{abstract}
\noindent 
Neutrino masses can be generated radiatively. In such scenarios their masses are calculated by evaluating a self-energy diagram with vanishing external  momentum, i.e.\ taking only the leading order term in a momentum expansion. 
The difference between the full self-energy and the mass is experimentally  difficult to access, since one needs off-shell neurinos to observe them. However, massive Majorana neutrinos that mediate neutrinoless double beta decay 
($0\nu\beta\beta$) are off-shell, with the virtuality of order 100 MeV. 
If the energy scale of the self-energy loop is of the order of this virtuality, the amplitude of double beta decay can be modified by  the unsuppressed loop effect. This can have a drastic impact on the interpretation of future observations of  the $0\nu\beta\beta$ decay.

\end{abstract}
\maketitle

\section{Introduction}
\noindent 
Neutrinoless double beta decay ($0\nu\beta\beta$) is a promising process  to probe the Majorana nature of neutrinos and the presence of lepton number violation in general: 
$$
(A,Z) \to (A,Z+2) + 2  e^- \,.
$$ 
A large number of experiments is currently runnning or under construction in order to observe or improve our limits on the decay \cite{Dolinski:2019nrj}. While the exchange of light massive Majorana masses is arguably the best-motivated mechanism  of the decay, countless other scenarios exist that can lead to $0\nu\beta\beta$, see e.g.\ \cite{Rodejohann:2011mu,Deppisch:2012nb,Graf:2018ozy} for reviews. 
In the standard neutrino mechanism of the decay, the value or limit 
of the effective mass that one extracts from the observed lifetime has 
to be compared to the neutrino mass observables in direct Kurie-plot searches and in cosmology. Consistency of the three complementary approaches would be a spectacular confirmation of the three-Majorana-neutrino paradigm. No consistency would be even more spectacular, as it would imply e.g.\ non-standard cosmology beyond $\Lambda$CDM or an alternative $0\nu\beta\beta$ mechanism. 

At the same time, the origin of neutrino mass is unknown. One option is that small neutrino masses are generated by loop-processes involving new particles beyond the Standard Model (SM), see e.g.\ 
Refs.\ \cite{Babu:2001ex,Ma:2009dk,Bonnet:2012kz,Sierra:2014rxa,Klein:2019iws} for systematic studies and \cite{Cai:2017jrq} for a recent review. In such radiative models neutrino mass is obtained by evaluating a  neutrino self-energy diagram and then setting the external momentum to zero. 

In this paper, we note that in radiative neutrino mass models the 
$0\nu\beta\beta$ decay rate could be enhanced by the neutrino self-energy 
loop. 
In $0\nu\beta\beta$ decay, massive neutrinos appear as intermediate states with a virtuality of ${\cal O}(100)$ MeV. 
Generally speaking, if the masses of the particles running in the neutrino-mass loop are below this internal 
momentum transfer of ${\cal O}(100)$ MeV, then the decay rate is sensitive to the full
self-energy diagram that generates neutrino mass\footnote{The presence of low-mass particles in radiative neutrino mass generation, 
though not often considered in the literature, is possible
and theoretically motivated in relation to small neutrino masses. 
}. 
Phenomenologically, the effect would be that the amplitude of the decay is modified by the new term coming from the self-energy diagram, which could  in the most straightforward scenario be a common enhancement or suppression of the amplitude. More complicated scenarios are also conceivable. 
In general, as for all non-standard mechanisms for $0\nu\beta\beta$, the interpretation of future limits or observations of the decay in  comparision with direct and cosmological neutrino mass approaches could change dramatically. \\


The paper is organized as follows. In Sec.~\ref{sec:basic} we first
briefly review the physical interpretation of neutrino self-energy
computed at the loop level and then adopt a model-independent approach
to study the influence of the radiative mass generation mechanism
on $0\nu\beta\beta$ decay. We discuss possible phenomenological  implications in Sec.~\ref{sec:result}. As an example for a realistic neutrino mass model, we apply in Sec.~\ref{sec:The-scotogenic-model} our 
conclusions to the scotogenic model. 
We conclude in Sec.~\ref{sec:Conclusion} and delegate technical details to an appendix.

\section{Model-independent study\label{sec:basic}}
\noindent 
In this section we study the influence of the radiative mass generation
mechanism on $0\nu\beta\beta$ decay in a model-independent approach.
The conclusions obtained in this section (see Tab.~\ref{tab:t}), 
in general, will apply to all 1-loop models up to ${\cal O}(1)$ factors.

First, let us briefly review the physical meaning of a loop-generated
neutrino mass. Consider the following tree-level Lagrangian of left-handed
neutrinos $\nu_{L}$:
\begin{equation}
{\cal L}_{{\rm tree}}=\overline{\nu_{L}}i\slashed{\partial}\nu_{L}-\frac{1}{2}\left[m_{0}\overline{\nu_{L}^{c}}\nu_{L}+{\rm h.c.}\right],\label{eq:-19}
\end{equation}
where $m_{0}$ is a tree-level Majorana neutrino mass. In the presence
of new neutrino interactions, there can be two types of 1PI (one-particle
irreducible) diagrams of neutrinos at loop level: (i) diagrams
with $\nu_{L}$ and $\overline{\nu_{L}}$ as external legs;  and (ii)
diagrams with $\nu_{L}$ and $\overline{\nu_{L}^{c}}$ as external
legs. We shall focus our discussion on the latter because the former
conserves the lepton number and only renormalizes the wavefunction.
If a 1PI diagram of type (ii) is evaluated, denoting the value as
$\Sigma(p^{\mu})$, where $p^{\mu}$ is the external neutrino momentum,
then this loop correction leads to the following
effective Lagrangian (in momentum space):
\begin{equation}
{\cal L}_{{\rm eff}}=\overline{\nu_{L}}\slashed{p}\nu_{L}-\frac{1}{2}\left(\overline{\nu_{L}^{c}}\left[m_{0}+\Sigma(p^{\mu})\right]\nu_{L}+{\rm h.c.}\right).\label{eq:-20}
\end{equation}
The effect of a constant $\Sigma(p^{\mu})$, or the possible zeroth order  term in an expansion in terms of 
$p^\mu$, can thus be identified as an additional contribution to the  neutrino mass. 
We can expand $\Sigma(p^{\mu})$ in terms of $p^{\mu}$: 
\begin{equation}
\Sigma(p^{\mu})=c_{0}+c_{1}p^{\mu}\gamma_{\mu}+c_{2}p^{\mu}p_{\mu}+\cdots\label{eq:-21}
\end{equation}
This is the only possible Lorentz invariant form the expansion can take. Checking the chirality, we can see that in $\overline{\nu_{L}^{c}}\left[m_{0}+\Sigma(p^{\mu})\right]\nu_{L}=\overline{\nu_{L}^{c}}P_{L}\left[m_{0}+\Sigma(p^\mu)\right]P_{L}\nu_{L}$, 
where $P_{L}=(1-\gamma^{5})/2$, the $c_{1}p^{\mu}\gamma_{\mu}$ term
vanishes because $P_{L}\gamma_{\mu}P_{L}=0$.  More generally, we
conclude that terms odd in $p^{\mu}$ vanish so that $\Sigma$ 
depends on $p^{2}\equiv p^{\mu}p_{\mu}$ only. It can thus  be written
as
\begin{equation}
\Sigma(p)=m_{\nu}\left[1+\frac{p^{2}}{\Lambda^{2}}+{\cal O}(p^{4})\right], \label{eq:master}
\end{equation}
where $m_\nu$ is the loop contribution to the neutrino mass, or, in absence of $m_0$, simply the neutrino mass. 
The scale $\Lambda$ corresponds, as we will demonstrate below, to masses of new particles participating in the loop mechanism. 
This expression implies that within radiative neutrino mass models the neutrino mass is obtained from a 
1PI diagram in the limit $p^\mu = 0$, i.e.\ for vanishing external momentum. 

\begin{figure}[t]
\centering

\includegraphics[width=0.8\textwidth]{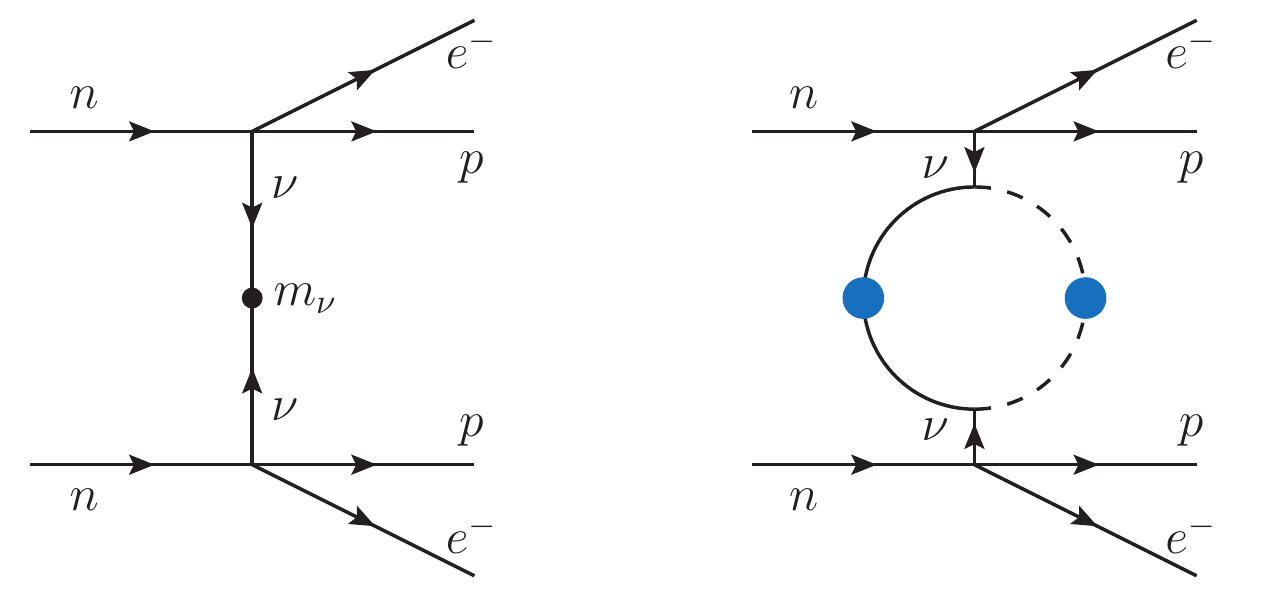}\caption{Left: Feynman diagram of the standard $0\nu\beta\beta$ decay process;
Right: Feynman diagram of a loop-enhanced $0\nu\beta\beta$ decay
process, where the blue blobs represent LNV, cf.\ 
Fig.\ \ref{fig:1PI}.
\label{fig:loop_0nbb}}
\end{figure}

In  neutrino oscillation experiments the intermediate propagating neutrino mass states are essentially immediately (as soon as they travel a distance $x \sim 1/p$) on-shell, i.e.\ $p^{2}\approx0$. In order to find a situation in which neutrinos are off-shell, or possess a large virtuality, we consider $0\nu\beta\beta$ decay. In this process, the exchanged Majorana neutrino (see Fig.\ \ref{fig:loop_0nbb}) is off-shell. In fact, it has a momentum corresponding to the distance of the two neutrons participating in $0\nu\beta\beta$: 
\[
p^{2}\sim\frac{1}{r^{2}}.
\]
The mean distance between neutrons for most heavy isotopes is\footnote{\label{fn:p-value}For more details, see, e.g., Fig.~1 in Ref.~\cite{Simkovic:2018hiq} or Fig.~5 in Ref.~\cite{Shimizu:2017qcy}.}:
\[
\langle r\rangle\approx1\ {\rm fm},
\]
corresponding to an energy scale of 200 MeV. Therefore if neutrino  
masses are generated by a loop with $\Lambda$ not much higher than
200 MeV, then the $p^{2}/\Lambda^{2}$ term will make a considerable
contribution to the $0\nu\beta\beta$ rate. In terms of diagrams, the  loop that generates neutrino mass radiatively appears in the internal neutrino line of the process. The decay rate should thus be computed using the right 
diagram in Fig.~\ref{fig:loop_0nbb} instead of
the left standard diagram. The amplitude of the standard
diagram is proportional to Majorana mass $m_{\nu}$. The loop effect
can be easily included by replacing the Majorana mass $m_{\nu}$ with
 $\Sigma$. However, the loop effect 
on the total $0\nu\beta\beta$ decay rate is in general more complicated, since it involves an integral over the 
momentum $p$ when the full calculation of the decay rate is performed. \\

In the above discussion, we do not take the flavor structure into
consideration. We need to generalize the neutrino self-energy $\Sigma$ and the mass $m_{\nu}$
to matrices with flavor indices ($\alpha$, $\beta$) added, such
as $\Sigma\rightarrow\Sigma_{\alpha\beta}$, $m_{\nu}\rightarrow(m_{\nu})_{\alpha\beta}$.
Since $\Lambda$ is defined in the $p$-expansion of $\Sigma$, it
has the same flavor indices as $\Sigma$.  In addition,
the neutrino self-energy may receive several loop contributions in
 realistic models, such as $\Sigma=\Sigma_{1}+\Sigma_{2}+\Sigma_{3}+\cdots$. An example, discussed later in Sec.\ \ref{sec:The-scotogenic-model}, is the scotogenic model in which three right-handed neutrinos are present, effectively generating neutrino mass by three 1-loop diagrams. 
Each contribution $\Sigma_{j}$ has a similar $p$-expansion
as Eq.~(\ref{eq:master}), with corresponding coefficients $m_{\nu j}$
and $\Lambda_{j}$. When summing these contributions together, one
can still use Eq.~(\ref{eq:master}) with the total neutrino mass matrix 
$m_{\nu}=m_{\nu1}+m_{\nu2}+m_{\nu3}+\cdots$ and $\Lambda^{2}$ given
by:
\begin{equation}
\frac{1}{\Lambda^{2}}=\sum_{j}\frac{1}{\Lambda_{j}^{2}}\frac{m_{\nu j}}{m_{\nu}}\,.\label{eq:Lambda_sum}
\end{equation}
Note that Eq.~(\ref{eq:Lambda_sum}) should be computed at the level of each matrix element. For instance, $\frac{1}{m_{\nu}}$ should be taken simply as 
$((m_{\nu})_{\alpha\beta})^{-1}$,  rather than the inverse of the matrix $m_{\nu}$ which would be $(m^{-1}_{\nu})_{\alpha\beta}$.

\begin{figure}
\centering

\includegraphics[width=0.8\textwidth]{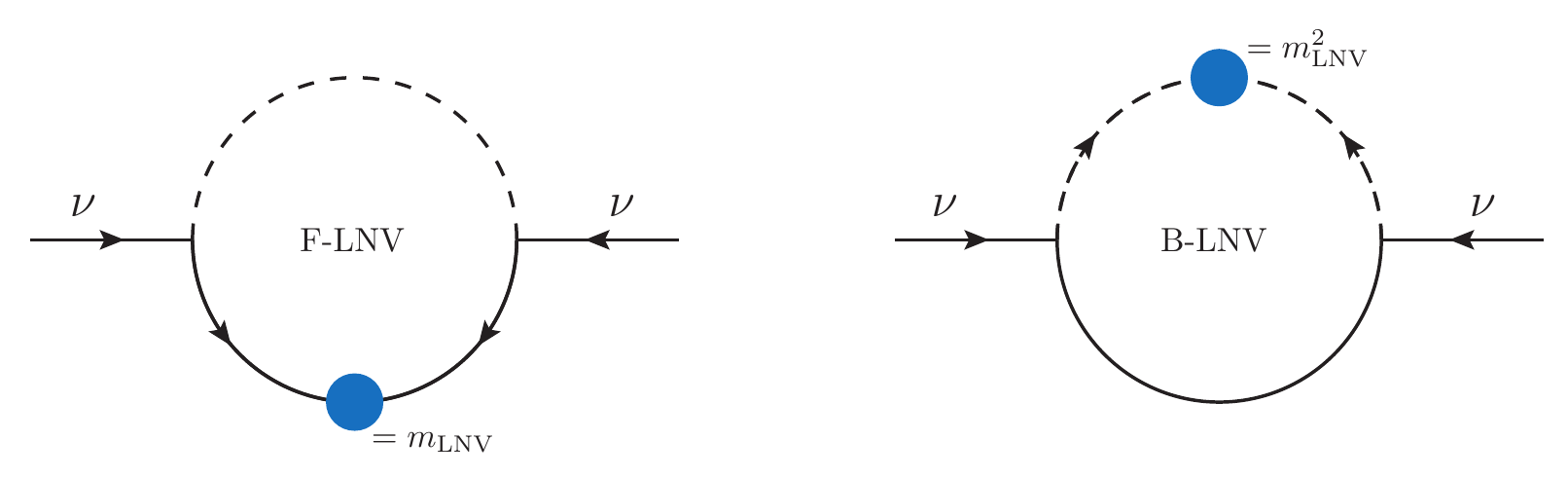}\caption{Lepton-number-violating neutrino self-energy  diagrams 
at the 1-loop level. To connect two external neutrino lines, the loop
can only consist of a bosonic line and a fermionic line. To generate 
Majorana neutrino masses, the loop must contain lepton number violation  (indicated by the
blue blobs). This can enter either in the fermionic part (left plot) 
or in the bosonic part (right) of the loop. Throughout this paper, 
we refer to the former as F-LNV and the latter as B-LNV.\label{fig:1PI}}
\end{figure}

Now let us discuss the generic topology of 1-loop diagrams that can
generate Majorana neutrino masses.  Since the loop must connect two
external fermion legs ($\nu_{L}$ and $\overline{\nu_{L}^{c}}$), it must consist of a fermion line and a boson line, as illustrated by the  diagrams in Fig.~\ref{fig:1PI}. 
This is the only possible topology for all 1-loop diagrams generating neutrino mass. The loops may have  additional scalar boson lines attached
(e.g., Fig.~\ref{fig:scoto}), which eventually end as vacuum expectation values (VEVs). Such 
lines can be removed in the context of computing the neutrino self-energy,
as their effects will be absorbed into the masses of particles running
in the loops. 
Next, generating Majorana neutrino masses requires sources of lepton number violation (LNV), which can appear either in the fermionic part or in the bosonic part of the loop,  see Fig.~\ref{fig:1PI}. 
We refer to the two
cases as Fermionic LNV (F-LNV) and Bosonic LNV (B-LNV), and discuss them  separately in what follows.

\subsection{F-LNV}

The left diagram in Fig.~\ref{fig:1PI} shows an F-LNV loop for radiative
neutrino mass generation. The LNV insertion (the blue blob) is essentially
a Majorana mass, though in complete models it may stand for more
complex structures that eventually give rise to such an effective
mass. Denoting the Majorana mass as $m_{\text{LNV}}$, the product
of the two Yukawa vertices as $y^{2}$, and the fermion/boson running
in the loop as $\psi$/$\phi$, respectively, we can compute this diagram
directly. Here we present the result and delegate the detailed calculation
 to Appendix~\ref{sec:Loop-calculation}.

The neutrino self-energy generated by the F-LNV loop is
\begin{eqnarray}
\Sigma & = & \frac{y^{2}}{16\pi^{2}}m_{\text{LNV}}\nonumber \\
 &  & \times\left[\frac{1}{\epsilon}+1-\frac{m_{\phi}^{2}\ln m_{\phi}^{2}-m_{\psi}^{2}\ln m_{\psi}^{2}}{m_{\phi}^{2}-m_{\psi}^{2}}+\frac{m_{\phi}^{4}-m_{\psi}^{4}-2m_{\psi}^{2}m_{\phi}^{2}\ln\frac{m_{\phi}^{2}}{m_{\psi}^{2}}}{2\left(m_{\phi}^{2}-m_{\psi}^{2}\right){}^{3}}p^{2}+{\cal O}(p^{4})\right],\label{eq:-0}
\end{eqnarray}
where $p$ is the neutrino momentum and $\epsilon=(4-d)/2$ is the commonly-used  notation in dimensional regularization. The masses of $\psi$ and
$\phi$ are denoted as $m_{\psi}$ and $m_{\phi}$, respectively. 

As can be expected, the result is proportional to $y^{2}$ 
and $m_{{\rm LNV}}$, with a typical loop-suppression factor  $(16\pi^{2})^{-1}$.
Putting $\frac{y^{2}}{16\pi^{2}}m_{\text{LNV}}$ aside, the remaining
part, though a little complicated, can be simply summarized as some
ratios of $m_{\psi}$, $m_{\phi}$ and $p^{2}$, independent of $m_{\text{LNV}}$.
The UV divergence $\frac{1}{\epsilon}$ is expected in the F-LNV loop,
but it will get cancelled in a complete renormalisable model ---see,
e.g., the scotogenic model discussed in Sec.~\ref{sec:The-scotogenic-model}. Note that from now on every time terms like $\ln m_x^2$ appear in  the expressions, it is understood that they actually read $\ln m_x^2/\mu^2$, with $\mu^2$ is the renormalization scale. Again, in complete renormalisable theories there is no dependence on $\mu^2$. 
In the presence of such complete models, $\frac{1}{\epsilon}$ is usually
replaced by a model-dependent ${\cal O}(1)$ quantity. 
When adopting a model-independent approach, it is usually safe to
disregard the $\frac{1}{\epsilon}$ part and just keep in mind that
the results may have ${\cal O}(1)$ uncertainties. See Refs.\ \cite{Bischer:2018zbd,Xu:2019dxe} for similar approaches.  

Note that in some models where $\psi$ is simply a right-handed neutrino,
$m_{\text{LNV}}$ and $m_{\psi}$ are identical. We prefer to have
two different masses here for more general consideration and also
for more manifest physical meaning of the expression. On the other
hand, the two fermion propagators connected by $m_{{\rm LNV}}$ could
be of different fermions with different masses, and each of the propagators
(including the one of the boson) could be further split into two or more propagators.
In this case, the result will have more complicated mass dependence\,---\,see
Eqs.~(\ref{eq:a-1}) and (\ref{eq:a-2}) for example. We refrain
from introducing more mass parameters here since two masses are sufficient
to illustrate the generic features of the F-LNV loop to be discussed
below.

Let us discuss interesting limits of Eq.~(\ref{eq:-0}) 
which can help us to further understand the dependence of the F-LNV
loop on the masses.

\vspace{10pt}

\noindent $\bullet$ Domination of $m_{\phi}$ ($m_{\psi}\ll m_{\phi}$): 

\vspace{5pt}
\noindent
When $m_{\psi}$ is much smaller than $m_{\phi}$, the result turns
out to be dominated by the contribution of $m_{\phi}$. In the limit
$m_{\psi}\rightarrow0$, Eq.~(\ref{eq:-0}) reduces to
\begin{equation}
\Sigma=\frac{y^{2}}{16\pi^{2}}m_{\text{LNV}}\left[\frac{1}{\epsilon}+1-\ln m_{\phi}^{2}+\frac{p^{2}}{2m_{\phi}^{2}}+{\cal O}(p^{4})\right].\label{eq:-1}
\end{equation}
For small nonzero $m_{\psi}$, the next-to-leading order (NLO) correction
to  the terms in the square bracket in Eq.~(\ref{eq:-1}) is:
\begin{equation}
\frac{m_{\psi}^{2}}{m_{\phi}^{2}}\ln\frac{m_{\psi}^{2}}{m_{\phi}^{2}}+\frac{m_{\psi}^{2}}{2m_{\phi}^{4}}\left(3+2\frac{m_{\psi}^{2}}{m_{\phi}^{2}}\ln\frac{m_{\psi}^{2}}{m_{\phi}^{2}}\right)p^{2}\,.\label{eq:-4}
\end{equation}

\vspace{10pt}

\noindent $\bullet$ Domination of $m_{\psi}$ ($m_{\psi}\gg m_{\phi}$): 

\vspace{5pt}
\noindent
Similarly, when $m_{\phi}$ is much smaller than $m_{\psi}$, the
result is not sensitive to $m_{\phi}$. Taking the limit $m_{\phi}\rightarrow0$,
Eq.~(\ref{eq:-0}) reduces to
\begin{equation}
\Sigma=\frac{y^{2}}{16\pi^{2}}m_{\text{LNV}}\left[\frac{1}{\epsilon}+1-\ln m_{\psi}^{2}+\frac{p^{2}}{2m_{\psi}^{2}}+{\cal O}(p^{4})\right].\label{eq:-2}
\end{equation}
The NLO correction of small nonzero $m_{\phi}$ to the terms in the
square bracket in Eq.~(\ref{eq:-2}) is
\begin{equation}
\frac{m_{\phi}^{2}}{m_{\psi}^{2}}\ln\frac{m_{\phi}^{2}}{m_{\psi}^{2}}+\frac{m_{\phi}^{2}}{2m_{\psi}^{4}}\left(3+2\frac{m_{\phi}^{2}}{m_{\psi}^{2}}\ln\frac{m_{\phi}^{2}}{m_{\psi}^{2}}\right)p^{2}\,.\label{eq:-5}
\end{equation}
Note that Eqs.~(\ref{eq:-2}) and (\ref{eq:-5}) can be obtained  by
interchanging $m_{\psi}\leftrightarrow m_{\phi}$ in Eqs.~(\ref{eq:-1})
and (\ref{eq:-4}), since Eq.~(\ref{eq:-0}) is symmetric under $m_{\psi}\leftrightarrow m_{\phi}$.

\vspace{10pt}

\noindent $\bullet$ The $m_{\psi}=m_{\phi}$ case:

\vspace{5pt}
\noindent
At first sight Eq.~(\ref{eq:-0}) seems to possess a singularity when 
$m_{\psi}$ approaches $m_{\phi}$. However, for 
 $m_{\psi}=m_{\phi}=m$, Eq.~(\ref{eq:-0}) reduces to:
\begin{equation}
\Sigma=\frac{y^{2}}{16\pi^{2}}m_{\text{LNV}}\left(\frac{1}{\epsilon}-\ln m^{2}+\frac{p^{2}}{6m^{2}}+{\cal O}(p^{4})\right).\label{eq:-3}
\end{equation}\ \\

From the above discussions, we can summarize that F-LNV loops typically
generate a result of the form given in Eq.~(\ref{eq:master}), where
the energy scale $\Lambda$ is mainly determined by the largest mass
in the loop:
\begin{equation}
\Lambda\approx{\cal O}\left[\max(m_{\psi},\ m_{\phi})\right].\label{eq:-18}
\end{equation}
We will show in the next subsection that B-LNV loops generate similar
result---for comparison, see Tab.~\ref{tab:t}.

\begin{table*}
\caption{\label{tab:t}Neutrino masses ($m_{\nu}$) and self-energies ($\Sigma$)
obtained from computing the 1-loop F-LNV and B-LNV diagrams in Fig.~\ref{fig:1PI}.}

\begin{ruledtabular}
\begin{tabular}{ccc}
 & $m_{\phi}$-dominated & $m_{\psi}$-dominated\tabularnewline
\hline 
\rule[-4ex]{0pt}{8ex} F-LNV  & $m_{\nu}=\frac{y^{2}}{16\pi^{2}}m_{{\rm LNV}}\times{\cal O}(1)$ & $m_{\nu}=\frac{y^{2}}{16\pi^{2}}m_{{\rm LNV}}\times{\cal O}(1)$ \tabularnewline
\rule[-4ex]{0pt}{8ex} & $\Sigma\approx m_{\nu}\left[1+\frac{p^{2}}{m_{\phi}^{2}}\times{\cal O}(1)\right]$ & $\Sigma\approx m_{\nu}\left[1+\frac{p^{2}}{m_{\psi}^{2}}\times{\cal O}(1)\right]$\tabularnewline
\hline 
\rule[-4ex]{0pt}{8ex} B-LNV & $m_{\nu}=\frac{y^{2}}{16\pi^{2}}m_{{\rm LNV}}^{2}\frac{m_{\psi}}{m_{\phi}^{2}}\times{\cal O}(1)$  & $m_{\nu}=\frac{y^{2}}{16\pi^{2}}\frac{m_{{\rm LNV}}^{2}}{m_{\psi}}\times{\cal O}(1)$\tabularnewline
\rule[-4ex]{0pt}{8ex} & $\Sigma\approx m_{\nu}\left[1+\frac{p^{2}}{m_{\phi}^{2}}\times{\cal O}(1)\right]$ & $\Sigma\approx m_{\nu}\left[1+\frac{p^{2}}{m_{\psi}^{2}}\times{\cal O}(1)\right]$\tabularnewline
\end{tabular}\end{ruledtabular}

\end{table*}

\subsection{B-LNV}

If neutrino masses are generated by a B-LNV loop shown in the right 
plot of Fig.~\ref{fig:1PI}, Majorana fermions are not 
required. Instead, the scalar boson in the loop has to
carry  lepton number and, typically via some VEV insertion, breaks it. 
The minimal UV-complete model that generates neutrino
masses via a B-LNV loop, to our knowledge,  is the Zee model \cite{Zee:1980ai}. 
In the Zee model, the fermion running in the loop is a SM charged
lepton, and the LNV part of the loop is achieved by a trilinear interaction of an $SU(2)_{L}$ singlet scalar scalar and two Higgs doublets. The latter obtain nonzero VEVs, which effectively give rise to the B-LNV structure discussed here---see Fig.~2 of Ref.~\cite{Zee:1980ai}. 

Note that the blue blob in the right panel of Fig.~\ref{fig:1PI}
has dimension of $[{\rm mass}]{}^{2}$, which we shall denote as $m_{\text{LNV}}^{2}$.
This is an important difference between the F-LNV and B-LNV loops.
In Appendix~\ref{sec:Loop-calculation}, we calculate the B-LNV loop.
 Using a similar notation as in the previous analysis on F-LNV,  the
calculation results in:
\begin{eqnarray}
\Sigma & = & \frac{y^{2}}{16\pi^{2}}m_{\text{LNV}}^{2}m_{\psi}\frac{m_{\psi}^{2}-m_{\phi}^{2}+m_{\psi}^{2}\ln\frac{m_{\phi}^{2}}{m_{\psi}^{2}}}{\left(m_{\phi}^{2}-m_{\psi}^{2}\right){}^{2}}\nonumber \\
 &  & \times\left[1-\frac{m_{\phi}^{4}+4m_{\psi}^{2}m_{\phi}^{2}-5m_{\psi}^{4}-2\left(m_{\psi}^{2}+2m_{\phi}^{2}\right)m_{\psi}^{2}\ln\frac{m_{\phi}^{2}}{m_{\psi}^{2}}}{2\left(m_{\phi}^{2}-m_{\psi}^{2}\right){}^{2}\left(m_{\psi}^{2}-m_{\phi}^{2}+m_{\psi}^{2}\ln\frac{m_{\phi}^{2}}{m_{\psi}^{2}}\right)}p^{2}+{\cal O}(p^{4})\right].\label{eq:-6}
\end{eqnarray}
Since the B-LNV loop contains two scalar mediators and one fermion
mediator, the loop integral is finite so there is no UV divergence
in Eq.~(\ref{eq:-6}). Similar to the previous discussion on F-LNV,
we can also derive some useful limits for B-LNV:

\vspace{10pt}

\noindent $\bullet$ Domination of $m_{\phi}$ ($m_{\psi}\ll m_{\phi}$):

\vspace{5pt}
\noindent
When $m_{\psi}\ll m_{\phi}$,  the denominators and numerators  
in Eq.~(\ref{eq:-6}) will be dominated by the highest powers of $m_{\phi}$, leading to:
\begin{equation}
\Sigma=-\frac{y^{2}}{16\pi^{2}}m_{\text{LNV}}^{2}m_{\psi}\left[\frac{1}{m_{\phi}^{2}}+\frac{p^{2}}{2m_{\phi}^{4}}+{\cal O}(p^{4})\right],\label{eq:-7}
\end{equation}
which implies that for large $m_{\phi}$, the neutrino self-energy
and mass are suppressed by $\frac{1}{m_{\phi}^{2}}$.

\vspace{10pt}

\noindent $\bullet$ Domination of $m_{\psi}$ ($m_{\psi}\gg m_{\phi}$):

\vspace{5pt}
\noindent
When $m_{\psi}\gg m_{\phi}$, the denominators and numerators 
in Eq.~(\ref{eq:-6}) will be dominated by the highest powers of $m_{\psi}$, leading to:
\begin{equation}
\Sigma=\frac{y^{2}}{16\pi^{2}}\frac{m_{\text{LNV}}^{2}}{m_{\psi}}\left[\ln\frac{m_{\phi}^{2}}{m_{\psi}^{2}}+1+\frac{2\ln\frac{m_{\phi}^{2}}{m_{\psi}^{2}}+5}{2m_{\psi}^{2}}p^{2}+{\cal O}(p^{4})\right],\label{eq:-8}
\end{equation}
which implies that for large $m_{\psi}$, the neutrino self-energy
and mass are suppressed by $\frac{1}{m_{\psi}}$.

\vspace{10pt}

\noindent $\bullet$ The $m_{\psi}=m_{\phi}$ case:

\vspace{5pt}
\noindent
After expanding Eq.~(\ref{eq:-6}) in terms of $m_{\psi}-m_{\phi}=\delta m$
and then taking the leading order, we get 
\begin{equation}
\Sigma=-\frac{y^{2}}{16\pi^{2}}m_{\text{LNV}}^{2}\frac{1}{m}\left[\frac{1}{2}+\frac{1}{12m^{2}}p^{2}+{\cal O}(p^{4})\right],\label{eq:-9}
\end{equation}
where $m\equiv m_{\psi}=m_{\phi}$. This implies that if $m_{\psi}$
and $m_{\phi}$ both are large and of the same order $m$, then
the neutrino self-energy and mass are suppressed by $\frac{1}{m}$.\\

Therefore, the neutrino self-energy generated by the B-LNV loop, in
the three cases discussed above, can also be summarized by Eq.~(\ref{eq:master})
where $\Lambda$ is, again, mainly determined by the largest mass
in the loop---see Eq.~(\ref{eq:-18}). However, the neutrino mass
$m_{\nu}$ in Eq.~(\ref{eq:master}) in the B-LNV case is generically
suppressed by the largest mass in the loop (see the summary in Tab.~\ref{tab:t}),
unlike the F-LNV case, where $m_{\nu}$ is typically proportional
to $\frac{y^{2}}{16\pi^{2}}$ and $m_{\text{LNV}}$ without further mass 
suppression\footnote{However, in concrete models, $m_{\text{LNV}}$ may have a more fundamental origin so that it is suppressed by other heavy particles involved.}. This is because the B-LNV and F-LNV loops are generically proportional to $m_{\text{LNV}}^{2}$ and $m_{\text{LNV}}$ respectively. Simply by dimensional arguments, the former needs to be attached with some quantity that has dimension of $[{\rm mass}]^{-1}$, which turns out to be $m_{\psi}/m_{\phi}^{2}$ in the $m_{\phi}$-dominated case, and $1/m_{\psi}$ in the $m_{\psi}$-dominated case.

\section{Phenomenology\label{sec:result}}

\noindent
In the standard scenario, the amplitude of $0\nu\beta\beta$ is  proportional to $ee$-element of the neutrino mass matrix, given by  
\begin{equation}
\langle m_{\beta\beta}\rangle=\left|m_{1}c_{12}^{2}c_{13}^{2}+e^{2i\alpha}m_{2}s_{12}^{2}c_{13}^{2}+e^{2i\beta}m_{3}s_{13}^{2}\right|,\label{eq:-17}
\end{equation}
where $s_{12}^{2}=1-c_{12}^{2}\approx0.297$; $s_{13}^{2}=1-c_{13}^{2}\approx2.14\times10^{-2}$;
$\alpha$ and $\beta$ are two unknown Majorana phases; the neutrino masses $m_{1}$, $m_{2}$ and $m_{3}$ can be determined from the two observed 
mass-squared differences, 
given the value of the lightest neutrino mass and the mass ordering (normal or inverted). 

\begin{figure}[h]
\centering
\includegraphics[width=0.85\textwidth]{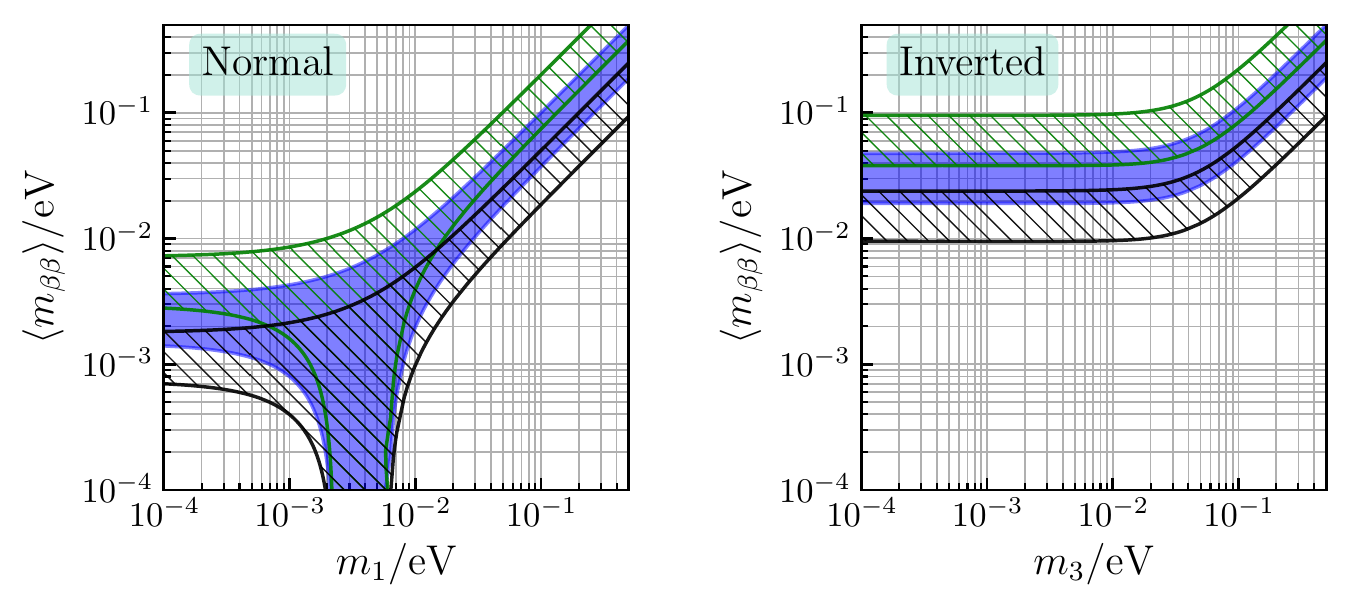}

\includegraphics[width=0.85\textwidth]{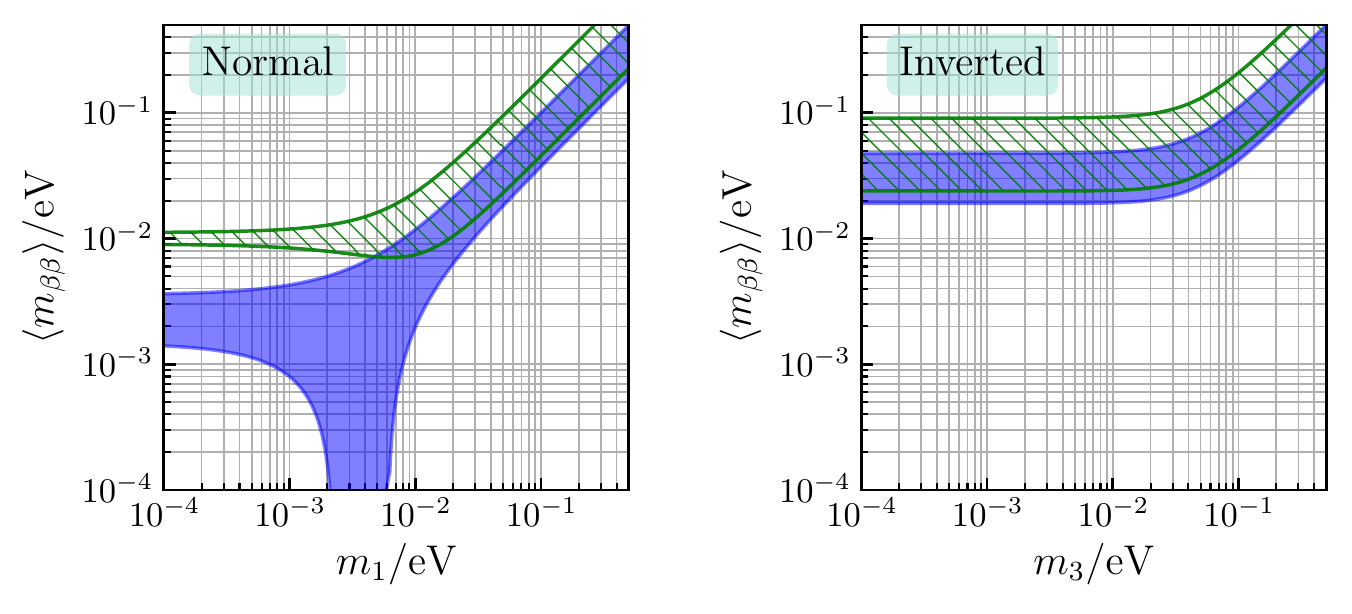}\caption{\label{fig:pheno}Loop-modified $0\nu\beta\beta$ decay. The blue
regions are the standard predictions of $\langle m_{\beta\beta}\rangle$
constructed from current best-fit values of oscillation parameters and  arbitrary Majorana phases. The upper plots assume a common suppression (black hatched) or enhancement (green hatched) of the effective mass by a factor of 2. The lower plots 
assume that $\Lambda_{2}^{2}=\langle p^{2}\rangle/3$ 
and negligible contributions from $\Lambda_{1}$ and $\Lambda_{3}$, see 
Eq.\ (\ref{eq:-22}). 
The left and right plots are for the normal and inverted mass ordering,
respectively.}
\end{figure}

As explained above, if there is a single 1-loop diagram generating  neutrino mass, and the involved particles are not much heavier than the neutrino's virtuality of ${\cal O}(100)$ MeV, we can replace the effective mass with (see the right diagram in Fig.\ \ref{fig:loop_0nbb})
\begin{equation}
\langle m_{\beta\beta}\rangle \to 
\langle m_{\beta\beta}\rangle \left(1 + \frac{p^2}{\Lambda^2} \right).
\end{equation}
This would enhance or supress the amplitude by a common ${\cal O}(1)$  factor, depending on the sign of the correction. Recall that the decay rate, as previously discussed, involves an integral over the neutrino momentum $p$, implying in an explicit model an additional relative  coefficient in front of $\frac{p^2}{\Lambda^2}$ that comes from this issue. We can ignore this complication, or alternatively assume that this factor can be hidden by redefining $\Lambda$.  
In Fig.\ \ref{fig:pheno} we show an example for the redefined effective mass that is a factor of 2 smaller or larger than the standard case.  Interesting things could happen. For instance, within the inverted ordering, one could interpret the measurement as an effective mass value lower than the usual minimal value in this case. One would assume now that another mechanism, probably of TeV-scale, generates the decay and that neutrinos are mainly Dirac or that the alternative TeV-scale mechanism interferes negatively with the standard neutrino diagram. However, it is more or less the usual diagram that mediates the decay, simple the self-energy term plays an important  role. A similar example is when, for quasi-degenerate neutrinos, the measurement would be interpreted as an effective mass value above the maximal allowed value for this case.

Within a loop mechanism for neutrino mass there can be several contributions to the neutrino mass matrix, 
see Eq.\ (\ref{eq:Lambda_sum}). This can be 
readily included into the $0\nu\beta\beta$
amplitude by the following replacement:
\begin{equation}
m_{i}\rightarrow m_{i}\left(1+\frac{p^{2}}{\Lambda_{i}^{2}}\right),\ (i=1,\ 2,\ 3),\label{eq:-22}
\end{equation}
where $\Lambda_{1}$, $\Lambda_{2}$, and $\Lambda_{3}$ in principle
can be different. When $\Lambda_{1}=\Lambda_{2}=\Lambda_{3}\equiv\Lambda$, we have the case discussed above, i.e.\ an overall enhancement or suppression of the amplitude. When 
$\Lambda_{1}$, $\Lambda_{2}$, and $\Lambda_{3}$ are different,
however, there can be more interesting phenomenology. For example, 
it has been well known that in the normal mass ordering, vanishing 
$\langle m_{\beta\beta}\rangle$ is possible due to cancellation among
the three terms in Eq.~(\ref{eq:-17}). If $m_{1}$, $m_{2}$ and 
$m_{3}$ are enhanced or suppressed differently, then such cancellation may be less complete 
or even disappear. As an illustrative example, in Fig.~\ref{fig:pheno} 
we show how $\langle m_{\beta\beta}\rangle$ can be changed if $\Lambda_{2}^{2}=\langle p^{2}\rangle/3$, 
while $\Lambda_{1}^{2}$ and $\Lambda_{2}^{2}$ are assumed to be
much larger than $\langle p^{2}\rangle$ so that their effects are
negligible. We can see for both the normal and inverted mass ordering that 
$\langle m_{\beta\beta}\rangle$ is enhanced from the blue standard regions 
to the new green regions. The most noteworthy change is that $\langle m_{\beta\beta}\rangle$ in the normal mass ordering cannot vanish anymore. 

\section{\label{sec:The-scotogenic-model}The scotogenic model as an example}
\noindent 
Now we shall apply our so far model-independent study to a specific example, by choosing the scotogenic model \cite{Ma:2006km}. 
By adding a new Higgs doublet and singlet fermions to
the SM, and introducing an unbroken $Z_{2}$ symmetry, the scotogenic model successfully accommodates a dark matter candidate and radiatively generates neutrino masses in a very economic way. 

Denoting the new Higgs doublet and the singlet fermions as $\eta=(\eta^{+},\ \eta^{0})^{T}$
and $N_j$, the model can be formulated as follows:
\begin{eqnarray}
{\cal L} &=&  {\cal L}_{{\rm {\rm SM}}}-y_{\alpha j}\overline{L_{\alpha}}\tilde{\eta}N_{j}-\frac{1}{2}M_{j}\overline{N_{j}^{c}}N_{j} 
 +|D\eta|^{2}-m_{\eta}^{2}\eta^{\dagger}\eta+\left({\rm terms\ of\ }\mbox{\ensuremath{\eta}}^{4}\ {\rm and}\ \eta^{2}H^{2}\right). \label{eq:-10}
\end{eqnarray}
Here $y_{\alpha j}$ is the Yukawa coupling matrix with a flavor
index $\alpha=e,\ \mu,\ \tau$ and a mass-eigenstate index  $j=1,\ 2,\ 3$;
$L$ is a left-handed lepton doublet, and $\tilde{\eta}=(\eta^{0},\ -\eta^{+})^{\dagger}$.
Without loss of generality, we have diagonalized the Majorana mass
matrix of the $N_j$ in Eq.~(\ref{eq:-10}) to its mass eigenvalues 
$M_{j}$. The SM Higgs doublet $H$ does not couple to neutrinos directly
due to the $Z_{2}$ symmetry: $N_{j}\rightarrow-N_{j}$, $\eta\rightarrow-\eta$. 
Therefore, the usual Dirac mass in the Type I seesaw is forbidden
and the left-handed neutrinos are massless at tree level.

\begin{figure}
\centering

\includegraphics{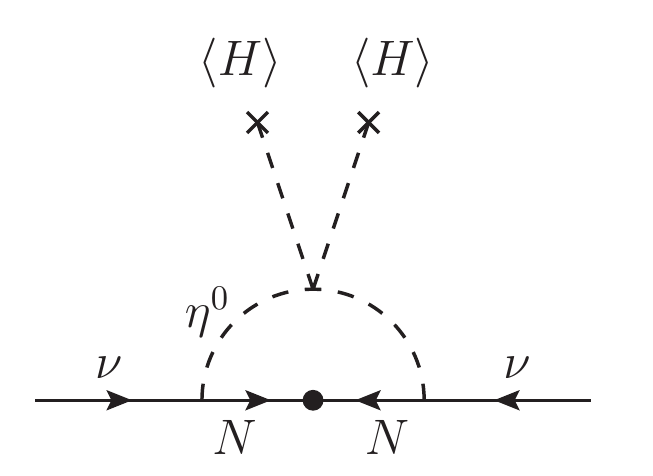}\caption{\label{fig:scoto}Feynman diagram for neutrino mass generation in
the scotogenic model. }
\end{figure}

The model generates neutrino masses at the 1-loop level via the diagram
presented in Fig.~\ref{fig:scoto}. As shown in this diagram, this 
is in our language an F-LNV loop.  It consists of $\eta^{0}$ (the
neutral component of $\eta$) and $N_j$ mediators, hence there are three different diagrams, corresponding to $N_{1,2,3}$. Their Majorana 
masses serve as the source of LNV. After electroweak symmetry
breaking, we replace the SM Higgs in this diagram by its VEV ($\eta$
does not obtain a VEV), which eventually contributes to the
masses of the scalar bosons running in the loop. Including the contribution of the SM Higgs VEV, the real and imaginary parts of $\eta^{0}$ become
mass eigenstates with different masses, denoted as $m_{R}$ and $m_{I}$, 
respectively. According to Ref.~\cite{Ma:2006km}, the neutrino
masses generated in this model are
\begin{equation}
\left(m_{\nu}\right)_{\alpha\beta}=\frac{y_{\alpha j}y_{\beta j}}{16\pi^{2}}M_{j}\left[\frac{m_{R}^{2}}{m_{R}^{2}-M_{j}^{2}}\ln\frac{m_{R}^{2}}{M_{j}^{2}}-\frac{m_{I}^{2}}{m_{I}^{2}-M_{j}^{2}}\ln\frac{m_{I}^{2}}{M_{j}^{2}}\right].\label{eq:-11}
\end{equation}
This result can also be quite straightforwardly obtained using our
model-independent calculation in Sec.~\ref{sec:basic}. From Eq.~(\ref{eq:-0}),
we can immediately write down the result in the case of ${\rm Re}(\eta^{0})$
running in the loop, simply by replacing $y^{2}\rightarrow y_{\alpha j}y_{\beta j}$,
$m_{{\rm LNV}}\rightarrow M_{j}$, $m_{\psi}\rightarrow M_{j}$, and
$m_{\phi}\rightarrow m_{R}$. The contribution
of the imaginary part ${\rm Im}(\eta^{0})$ is similar but differs
by a minus sign because the Yukawa coupling of ${\rm Im}(\eta^{0})$
is attached with an additional $i$: $y^{2}\rightarrow(iy_{\alpha j})(iy_{\beta j})=-y_{\alpha j}y_{\beta j}$.
Due to this minus sign, the terms $\frac{1}{\epsilon}+1$ in Eq.~(\ref{eq:-0}) from diagrams containing the real and imaginary parts of  $\eta^0$ cancel each other. The neutrino 
self-energy after this cancellation can be written as follows:
\begin{equation}
\Sigma_{\alpha\beta}=\left(m_{\nu}\right)_{\alpha\beta}\left[1+\frac{p^{2}}{\Lambda^{2}}+{\cal O}(p^{4})\right],\label{eq:-12}
\end{equation}
where
\begin{equation}
\Lambda^{2}=2\frac{\frac{m_{R}^{2}}{m_{R}^{2}-M_{j}^{2}}\ln\frac{m_{R}^{2}}{M_{j}^{2}}-\frac{m_{I}^{2}}{m_{I}^{2}-M_{j}^{2}}\ln\frac{m_{I}^{2}}{M_{j}^{2}}}{\frac{M_{j}^{4}-m_{R}^{4}+2m_{R}^{2}M_{j}^{2}\ln\frac{m_{R}^{2}}{M_{j}^{2}}}{\left(m_{R}^{2}-M_{j}^{2}\right)^{3}}-\frac{M_{j}^{4}-m_{I}^{4}+2m_{I}^{2}M_{j}^{2}\ln\frac{m_{I}^{2}}{M_{j}^{2}}}{\left(m_{I}^{2}-M_{j}^{2}\right)^{3}}}\,.\label{eq:-13}
\end{equation}
Eq.~(\ref{eq:-13}) applies only to a single $M_{j}$ contribution,
while to include multiple  contributions 
one needs to use Eq.~(\ref{eq:Lambda_sum}). 

The expression (\ref{eq:-13}) is rather complicated. In what follows,
we would like to focus on one hierarchical scenario ($M_{j}\ll m_{R}\ll m_{I}$)
to discuss the phenomenology. First, for $m_{\psi}\ll m_{R,I}$, the expansion in $M_j$ simplifies Eq.~(\ref{eq:-13})
to 
\begin{equation}
\Lambda^{2}=\frac{2m_{I}^{2}m_{R}^{2}}{m_{I}^{2}-m_{R}^{2}}\ln\frac{m_{I}^{2}}{m_{R}^{2}}+{\cal O}(M_{j}^{2}).\label{eq:-14}
\end{equation}
Then, assuming $m_{R}\ll m_{I}$, Eq.~(\ref{eq:-14}) further 
simplifies to 
\begin{equation}
\Lambda^{2}\approx2m_{R}^{2}\ln\frac{m_{I}^{2}}{m_{R}^{2}},\ {\rm for}\ M_{j}\ll m_{R}\ll m_{I},\label{eq:-16}
\end{equation}
which implies that  the energy scale of the loop in this scenario
is mainly determined by $m_{R}$, as long as the hierarchy $M_{j}\ll m_{R}\ll m_{I}$
holds.

Now we need to know how small $m_{R}$ and $M_{j}$ can be in this
model. 
First, let us inspect the singlet fermion masses. If $M_{j}$ is too
light, it would not generate the correct values of neutrino masses,
which should be around $0.05$ to $0.1$ eV. So without any particular
cancellation, Eq.~(\ref{eq:-11}) implies that $M_{j}$ needs to be
above $16\pi^{2} \times m_{\nu}={\cal O}(10)$ eV, assuming that the Yukawa
couplings cannot be much larger than ${\cal O}(1)$. For smaller Yukawa
couplings, $M_{j}$ correspondingly needs to be higher. On the other
hand, if the singlet fermions are lighter than a few MeV with sizable
coupling, they contribute to the effective number of neutrinos in the
early universe, excluded by BBN and CMB observations.
Therefore, to evade such bounds, we assume that $M_{j}\gtrsim10$
MeV. With this assumption, we get $y\sim4\pi\sqrt{m_{\nu}/M_{j}}\lesssim10^{-3}$, which is below meson decay bounds \cite{Barger:1981vd,Lessa:2007up,Pasquini:2015fjv} on neutrino-scalar couplings. 
Note that 
the neutral $\eta^{0}$ does not couple to charged leptons or quarks and can easily evade observational constraints. 
The scalar masses $m_{R}$ and $m_{I}$ cannot be both very light
otherwise the SM $Z$ boson would decay to ${\rm Re}(\eta^{0})$
and ${\rm Im}(\eta^{0})$. 
Therefore we only consider the case that $m_{I}$ is 
well above the $Z$ mass while $m_{R}$ is much lighter. Other 
collider constraints on $\eta$ include charged Higgs searches at
LEP and LEP II, electroweak precision test, and Higgs invisible
decay---see Refs.~\cite{Lundstrom:2008ai,Dolle:2009fn} for a detailed 
discussion. Charged Higgs searches have set a lower bound on the mass 
of $\eta^{\pm}$ in the form of $m_{\pm}\gtrsim70$--$90$ GeV \cite{Pierce:2007ut}. 
This combined with electroweak precision tests puts a similar
bound on $m_{I}$ because the $T$-parameter requires that $(m_{\pm}-m_{I})(m_{\pm}-m_{R})$ must be small \cite{Lundstrom:2008ai}. To obey both  constraints, we set $m_{\pm}=m_{I}\gtrsim100$ GeV. Higgs invisible decay 
could provide potentially important constraints since the SM Higgs 
could decay to two ${\rm Re}(\eta^{0})$ particles if their mass $m_{R}$ is
light. However, the current Higgs data from the LHC still allows for
about 10\% to 20\% invisible decay width~\cite{Aaboud:2018sfi,Sirunyan:2018owy}. 
On the other hand, the invisible decay width in this model can be
suppressed by tuning the quartic couplings in the scalar potential 
while still keeping the above scenario viable \cite{Barbieri:2006dq}. Such tuning is a general feature when one wants to be in the interesting situation in which the full 1PI diagram of the radiative neutrino mass mechanism is experimentally accessible for $0\nu\beta\beta$.

Nevertheless, we conclude that in the scotogenic model, at least the
parameter space with $m_{R}\gtrsim M_{j}\gtrsim10$ MeV and $m_{\pm}=m_{I}\gtrsim100$
GeV is allowed by various constraints. Within this parameter space,
$\Lambda$ in Eq.~(\ref{eq:-13}) can reach any value above the following
bound:
\begin{equation}
\Lambda\gtrsim111\ {\rm MeV}.\label{eq:-24}
\end{equation}
Taking $m_{R}=30$ MeV, $M_{j}=10$ MeV and $m_{\pm}=m_{I}=200$ GeV
for example (corresponding to $m_\nu \sim y^2$ MeV, which reproduces the typical neutrino mass scale for permille-level Yukawa couplings), Eq.~(\ref{eq:-16}) gives $178$ MeV while the full expression~(\ref{eq:-13}) gives $\Lambda\approx209\ {\rm MeV}$. This implies that $\Lambda$ in the scotogenic model can indeed be of the order of ${\cal O}(100)$ MeV.

\section{Conclusion\label{sec:Conclusion}}
\noindent
Radiative neutrino mass mechansims generate neutrino mass via a self-energy diagram in which one sets the external momentum to zero. 
The difference between the mass and the self-energy is usually not accessible in   neutrino oscillation experiments as long as the neutrinos are on-shell.
An interesting exception is neutrinoless double beta decay for which the neutrino virtuality is of order 100 MeV. Accepting the possibility that the particles in the neutrino-mass loop have masses around this scale, implies that higher order or even the full self-energy diagram have an effect in the decay. We performed a general study on the form of the diagram, and gave possible phenomenological consequences of a modified double beta decay amplitude. The scotogenic model was used as an explicit example to demonstrate that low-mass particles are possible in realistic models.

One can of course generalise the analysis to 2-, 3-, $n$-loop mechanisms for neutrino mass. 
We should note that there are radiative mechansims for Dirac neutrinos, which do not offer the option to see effects of the self-energy diagram if they do not contain LNV.


\begin{acknowledgments}\noindent
We thank Evgeny Akhmedov, Ernest Ma, Fedor Simkovic and Carlos Yaguna for useful discussions. WR was supported by the DFG with grant RO 2516/7-1 in the Heisenberg program. 
\end{acknowledgments}

\appendix

\section{Loop calculation of the F-LNV and B-LNV diagrams\label{sec:Loop-calculation}}
\noindent
Since the calculation involves charge conjugations of Dirac spinors,
we would like to review a few identities which will be used. First,
the notation $\psi^{c}$ of the general Dirac spinor $\psi$ is defined
as $\psi^{c}=-i\gamma^{2}\psi^{*}$, so we have 
\begin{equation}
\psi^{c}=\gamma^{c}\psi^{*},\ \overline{\psi^{c}}=\left(\gamma^{c}\psi^{*}\right)^{\dagger}\gamma^{0}=\left(\gamma^{0}\gamma^{c}\psi\right)^{T}\,,\label{eq:app}
\end{equation}
where we have introduced $\gamma^{c}\equiv-i\gamma^{2}$ for simplicity.
We are working with the chiral representations of Dirac matrices
so that
\begin{equation}
\gamma^{0}=\left(\gamma^{0}\right)^{T}=\left(\gamma^{0}\right)^{\dagger},\ \gamma^{c}=\left(\gamma^{c}\right)^{T}=\left(\gamma^{c}\right)^{\dagger},\ \left(\gamma^{0}\gamma^{c}\right)^{T}=-\gamma^{0}\gamma^{c}\,.\label{eq:app-1}
\end{equation}
In this convention, the transpose of $\gamma^{\mu}$ can be written
as
\begin{equation}
\left(\gamma^{\mu}\right)^{T}=\gamma^{0}\gamma^{c}\gamma^{\mu}\gamma^{0}\gamma^{c}\,,\label{eq:app-2}
\end{equation}
which can be useful when computing the transpose of a Dirac propagator.
With the above notations and identities, we can convert the following
Dirac propagators to each other: 
\[
\Delta_{\psi}\equiv\langle\psi\overline{\psi}\rangle\,,\ \Delta_{\psi}^{c}\equiv\langle\overline{\psi^{c}}\,\overline{\psi}\rangle\,.
\]
Compared to the well-known propagator $\Delta_{\psi}$, the second
propagator $\Delta_{\psi}^{c}$ is given less often, so here
we derive it briefly. Let us denote the Dirac indices (using $e$,
$f$, $g$, $\cdots$) explicitly, then we have 
\begin{equation}
\left[\Delta_{\psi}^{c}\right]_{ef}\equiv\langle\overline{\psi_{e}^{c}}\,\overline{\psi_{f}}\rangle=\langle\left(\gamma^{0}\gamma^{c}\psi\right)_{e}\overline{\psi_{f}}\rangle=\sum_{g}\langle\left(\gamma^{0}\gamma^{c}\right)_{eg}\psi_{g}\overline{\psi_{f}}\rangle=\sum_{g}\left(\gamma^{0}\gamma^{c}\right)_{eg}\left[\Delta_{\psi}\right]_{gf}=\gamma^{0}\gamma^{c}\Delta_{\psi}\,.\label{eq:app-3}
\end{equation}
Denoting the mass and momentum of $\psi$  as $m_{\psi}$ and $q$
respectively, the explicit forms of $\Delta_{\psi}$, $\Delta_{\psi}^{c}$
and their transpose are 
\begin{equation}
\Delta_{\psi}=\frac{i(\slashed{q}+m_{\psi})}{q^{2}-m_{\psi}^{2}}\,,\ \Delta_{\psi}^{T}=\frac{i(\gamma^{0}\gamma^{c}\slashed{q}\gamma^{0}\gamma^{c}+m_{\psi})}{q^{2}-m_{\psi}^{2}}\,,\label{eq:app-4}
\end{equation}
\begin{equation}
\Delta_{\psi}^{c}=\gamma^{0}\gamma^{c}\frac{i(\slashed{q}+m_{\psi})}{q^{2}-m_{\psi}^{2}}\,,\ \left(\Delta_{\psi}^{c}\right)^{T}=\frac{i(\gamma^{0}\gamma^{c}\slashed{q}-m_{\psi})}{q^{2}-m_{\psi}^{2}}\,.\label{eq:app-5}
\end{equation}
 With these identities, let us compute the F-LNV and B-LNV loops.

\vspace{10pt}

\noindent $\bullet$ F-LNV:

\vspace{5pt}

\noindent
For generality, we consider that the two fermion propagators are of
two different fermions, denoted as $\psi$ and $\psi'$. The relevant
part of the Lagrangian is formulated as 
\begin{equation}
{\cal L}\supset y\phi\overline{\psi}\nu_{L}+y'\phi\overline{\psi'}\nu_{L}+m_{{\rm LNV}}\overline{\psi'^{c}}\psi\,.\label{eq:app-6}
\end{equation}
Integrating out the $\phi$, $\psi$ and $\psi'$ fields in Eq.~(\ref{eq:app-6})
will lead to an effective operator of two neutrinos
\begin{equation}
\overline{\nu_{L}^{c}}\Sigma\nu_{L}=-\left(\gamma^{0}\gamma^{c}\Sigma\right){}_{fg}\left(\nu_{L}\right)_{f}\left(\nu_{L}\right)_{g}\,,\label{eq:ap}
\end{equation}
which can be computed using the F-LNV Feynman diagram in Fig.~\ref{fig:1PI}
and Eqs.~(\ref{eq:app-4}) and (\ref{eq:app-5}): 
\begin{eqnarray}
-i\overline{\nu_{L}^{c}}\Sigma\nu_{L} & =- & yy'\int\frac{dq^{4}}{(2\pi)^{4}}\overline{\nu_{L}^{c}}P_{L}\frac{i(\slashed{q}+m_{\psi})}{q^{2}-m_{\psi}^{2}}\left(-im_{{\rm LNV}}\right)\frac{i(\slashed{q}-m_{\psi'})}{q^{2}-m_{\psi'}^{2}}\frac{i}{(q-p)^{2}-m_{\phi}^{2}}P_{L}\nu_{L}\nonumber \\
 & = & \frac{yy'}{16\pi^{2}}m_{{\rm LNV}}\left[I^{(0)}+I^{(2)}p^{2}+{\cal O}(p^{4})\right]\overline{\nu_{L}^{c}}\nu_{L}\,,\label{eq:a-0}
\end{eqnarray}
where we have used {\tt Package-X} \cite{Patel:2015tea} to evaluate
the loop integral and expanded the result in terms of $p^{2}$ with
$I^{(0)}$ and $I^{(2)}$ given by: 
\begin{eqnarray}
I^{(0)} & = & \frac{1}{\epsilon}+1-\ln m_{\psi}^{2}+\frac{m_{\phi}^{2}m_{\psi}+m_{\phi}^{2}m_{\psi'}-m_{\psi}^{2}m_{\psi'}}{\left(m_{\psi}^{2}-m_{\phi}^{2}\right)\left(m_{\psi}+m_{\psi'}\right)}\ln\frac{m_{\phi}^{2}}{m_{\psi}^{2}}\nonumber \\
 &  & +\frac{m_{\psi'}^{3}}{\left(m_{\psi'}^{2}-m_{\phi}^{2}\right)\left(m_{\psi}+m_{\psi'}\right)}\ln\frac{m_{\phi}^{2}}{m_{\psi'}^{2}}\,,\label{eq:a-1}
\end{eqnarray}
\begin{eqnarray}
I^{(2)} & = & \frac{\left(m_{\phi}^{2}+m_{\psi}m_{\psi'}\right)\left(m_{\phi}^{4}+m_{\phi}^{2}m_{\psi}^{2}+m_{\phi}^{2}m_{\psi'}^{2}+m_{\psi}^{2}m_{\psi'}^{2}-4m_{\phi}^{2}m_{\psi}m_{\psi'}\right)}{2\left(m_{\phi}^{2}-m_{\psi}^{2}\right)^{2}\left(m_{\phi}^{2}-m_{\psi'}^{2}\right)^{2}}+\nonumber \\
 &  & +\frac{m_{\phi}^{2}}{m_{\psi}+m_{\psi'}}\left(\frac{m_{\psi}^{3}}{\left(m_{\psi}^{2}-m_{\phi}^{2}\right)^{3}}\ln\frac{m_{\phi}^{2}}{m_{\psi}^{2}}+\frac{m_{\psi'}^{3}}{\left(m_{\psi'}^{2}-m_{\phi}^{2}\right)^{3}}\ln\frac{m_{\phi}^{2}}{m_{\psi'}^{2}}\right).\label{eq:a-2}
\end{eqnarray}
Taking the $m_{\psi'}\rightarrow m_{\psi}$ limit, one can straightforwardly
obtain the result in Eq.~(\ref{eq:-0}).

\vspace{10pt}

\noindent $\bullet$ B-LNV:

\vspace{5pt}
\noindent
Similar to the F-LNV case, we also consider that the two scalar propagators
are of two different scalars, denoted as $\phi$ and $\phi'$. The
relevant part of the Lagrangian is formulated as 
\begin{equation}
{\cal L}\supset y\phi\overline{\psi^{c}}\nu_{L}+y'\phi'\thinspace\overline{\psi}\nu_{L}+m_{{\rm LNV}}^{2}\phi'^{*}\phi \,.\label{eq:app-6-1}
\end{equation}
Integrating out the $\psi$, $\phi$ and $\phi'$ gives
\begin{eqnarray}
-i\overline{\nu_{L}^{c}}\Sigma\nu_{L} & =- & y^{2}\int\frac{dq^{4}}{(2\pi)^{4}}\overline{\nu_{L}^{c}}\frac{i(\slashed{q}-m_{\psi})}{q^{2}-m_{\psi}^{2}}\frac{i}{(q-p)^{2}-m_{\phi}^{2}}\left(-im_{{\rm LNV}}^{2}\right)\frac{i}{(q-p)^{2}-m_{\phi'}^{2}}\nu_{L}\nonumber \\
 & = & -\frac{y^{2}}{16\pi^{2}}m_{{\rm LNV}}^{2}m_{\psi}\left[I^{(0)}+I^{(2)}p^{2}+{\cal O}(p^{4})\right]\overline{\nu_{L}^{c}}\nu_{L}\,,\label{eq:app-7}
\end{eqnarray}
where the result has been expanded in terms of $p^{2}$ with $I^{(0)}$
and $I^{(2)}$ given by:
\begin{eqnarray}
I^{(0)} & = & \frac{1}{m_{\phi}^{2}-m_{\phi'}^{2}}\left(\frac{m_{\phi'}^{2}}{m_{\phi'}^{2}-m_{\psi}^{2}}\ln\frac{m_{\phi'}^{2}}{m_{\psi}^{2}}-\frac{m_{\phi}^{2}}{m_{\phi}^{2}-m_{\psi}^{2}}\ln\frac{m_{\phi}^{2}}{m_{\psi}^{2}}\right),\label{eq:app-8}
\end{eqnarray}
\begin{eqnarray}
I^{(2)} & = & \frac{m_{\psi}^{2}}{m_{\phi}^{2}-m_{\phi'}^{2}}\left(\frac{m_{\phi'}^{2}}{\left(m_{\phi'}^{2}-m_{\psi}^{2}\right)^{3}}\ln\frac{m_{\phi'}^{2}}{m_{\psi}^{2}}-\frac{m_{\phi}^{2}}{\left(m_{\phi}^{2}-m_{\psi}^{2}\right)^{3}}\ln\frac{m_{\phi}^{2}}{m_{\psi}^{2}}\right)\nonumber \\
 &  & -\frac{\left(m_{\phi'}^{2}+m_{\phi}^{2}\right)m_{\psi}^{2}-3m_{\psi}^{4}+m_{\phi}^{2}m_{\phi'}^{2}}{2\left(m_{\phi}^{2}-m_{\psi}^{2}\right)^{2}\left(m_{\phi'}^{2}-m_{\psi}^{2}\right)^{2}}\,.\label{eq:app-9}
\end{eqnarray}
Taking the $m_{\phi'}\rightarrow m_{\phi}$ limit, one can straightforwardly
obtain the result in Eq.~(\ref{eq:-6}).

\bibliographystyle{JHEP}
\bibliography{ref}

\end{document}